\documentclass[%
 reprint,
 amsmath,amssymb,
 aps,
]{revtex4-2}

\usepackage{graphicx}
\usepackage{dcolumn}
\usepackage{bm}

\begin{document}

\preprint{APS/123-QED}

\title{Nanosecond kinetics of phase separation in binary fluid studied by a combination of Laser-induced temperature jump and Forster Resonant Excitation Transfer probing.} 

\author{Alexei Goun}
 
 \email{agoun@princeton.edu}
\affiliation{Department of Chemistry, Princeton University}

\date{\today}

\begin{abstract}
A novel technique for studying the demixing of binary fluids with nanosecond time resolution at nanometer spatial resolution is demonstrated. Nanosecond kinetics of phase separation initial stages in the water/2,6-lutidine mixture studied through the combination of a laser-induced temperature jump, and Forster Resonant Excitation Transfer techniques. A pair of chromophores: 8-hydroxy-1,3,6-pyrenetrisulfonate (HPTS) and Coumarin 480 form a FRET pair in the uniform phase of the mixture and separate into the components of the mixture in the nonuniform phase. The temperature jump that brings the mixture into a nonuniform phase is performed by the mid-infrared laser pulse. The exponential growth stage of phase separation is in the 5 to 20 nanoseconds range. The demonstrated technique can be applied to broad range of chemical and biological systems.

\end{abstract}

\maketitle

Despite the progress in understanding the near-equilibrium properties of systems undergoing the phase transition, a complete picture of the kinetics is still lacking. It is especially the case in the strongly non-equilibrium conditions of large temperature and pressure variations that are at the core of a large array of industrial, biological, and environmental processes \cite{rosenzweig2017eukaryotic}. Thus it is desirable to extend experimental and theoretical approaches into rapid, strongly nonequilibrium stages of phase transformations. 

A complete picture of the phase transition process would include measurements of the spatial organization on all time and space scales from molecular to macroscopic. The time resolution of phase separation measurements is often limited due to the finite heat transfer rate and thus the rate of cooling. The spatial resolution is often limited by either the resolving power of the microscope in imaging experiments or the wavelength of light in scattering experiments. Polymer systems \cite{bates1989spinodal,tran1997soft} have been thoroughly studied due to the characteristic time scale of spinodal decomposition being long. The dynamics of phase separation for simple binary mixtures is significantly faster \cite{mallamace1995spinodal, kubota1992spinodal,kubota1992spinodal_PRA,chou1979phase,huang1974study} and usually measured for very small changes of the temperature into the phase separating region with time scales of the measurement are from the late ms to late minutes time scale. The time scale of the initial phase separation process of either nucleation or spinodal decomposition is still out of reach for large, rapid changes in thermodynamic conditions. Molecular dynamics simulations show that phase separation can be rather rapid down to the nanosecond time scale given that the initial state is highly unequilibrated with a substantial amount of free energy. Hence, it is important to place a bound on the time scale of the initial stage of the phase separation. The experimental technique needs to combine high spatial and temporal resolution. To achieve high time resolution, a way to rapidly initiate the phase transition is required.
	Techniques that are most widely used to study phase transitions are either scattering techniques or direct imagining techniques. Direct imaging techniques give us the most informative route to study the kinetics and morphology of the phase organization. The spatial resolution can not be smaller than half the wavelength of light due to the Abbe limit, and separated phases have to be of sufficiently different refractive indices to provide an image with reasonable contrast. The scattering technique shares the wavelength of light limitations with the direct imaging techniques due to the unfavorable dependence of scattering intensity on the size of refractive index inhomogeneity \cite{pekker2017initial}. Thus, studying rapid, short spatial scale stages of phase transitions is impossible with currently available optical techniques.

\begin{figure}[b]
\includegraphics[width=0.5\textwidth]{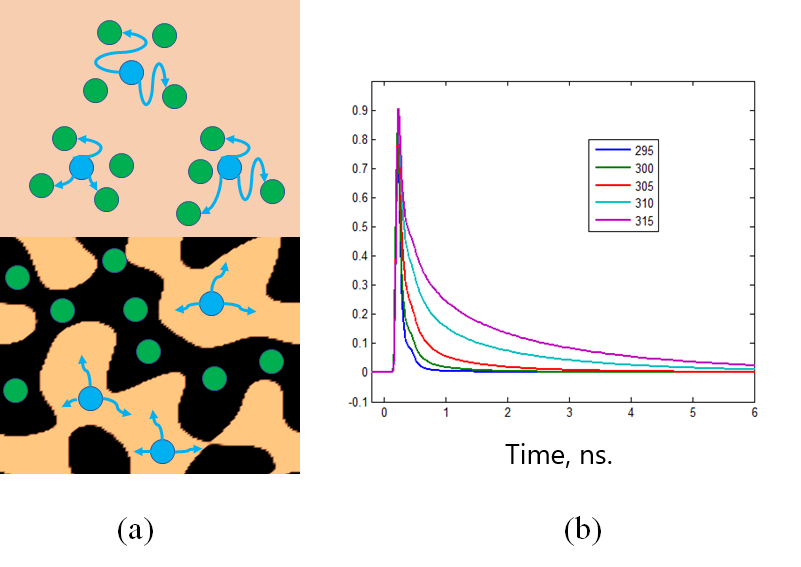}
\caption{\label{fig:id_expt} Left panel, The distribution of the donor/acceptor dye molecules in the uniform phase (top), (bottom) the distribution of the dye molecules when phases separated. Right panel, the kinetics of the donor (Coumarin 480) fluorescence decay  at various temperatures of the mixture, measured at the acceptor concentration of 50mM.}
\end{figure}

 In the present experiment, the Forster resonant excitation transfer (FRET) process is employed to elucidate the phase separation space scales down to tens of angstroms – tens of nanometers, inaccessible for direct optical imaging and scattering techniques.	The technique is based on the fluorescent excitation transfer between carefully selected dye molecules with different solubilities in the components of the mixture and is illustrated in Figure \ref{fig:id_expt}. Let us consider a binary mixture in the uniform phase, with a pair of distinct dye molecules with appropriate donor fluorescent/acceptor absorption spectra. The emission of the donor molecules overlaps with the absorption of the acceptor molecule. The Donor molecule will undergo an excitation transfer to the acceptor molecule once the acceptor concentration is high enough, Figure \ref{fig:id_expt}(a), upper panel. 

  Once thermodynamic conditions are changed, the mixture develops a spatially dependent composition distribution. Due to the different solubilities of the donor and acceptor molecules in the component of the mixture, each dye molecule would choose its own component to go to.

	\begin{figure}[b]

\includegraphics[width=0.5\textwidth]{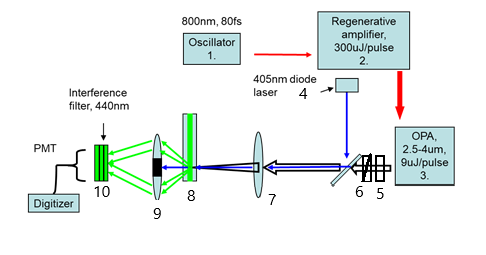}
\caption{\label{fig:exp_sys} Schematic of the experimental apparatus: 1 - femtosecond oscillator, 2 - regenerative amplifier, 3 - mid-IR Optical Parametric Amplifier,  4 - continuous wave diode laser (405 nm), 5 - mid-IR half-wave plate, 6 - mid-IR polarizer, 7 - focusing lens, 8 - spinning sample, 9 - fluorescence collecting lens, 10 - bandpass filter}
\end{figure}

 In the vicinity of the donor molecule, the concentration of the acceptor molecule would decrease, due to local changes in the composition. At the same time, the phase boundary is going to form at a certain distance from the donor molecule, consequently, the average separation of the donor-acceptor molecules is going to be influenced by the location of the phase boundary Figure \ref{fig:id_expt}(a) lower panel.

We have identified two molecules as a suitable donor/acceptor pair. Coumarine480 serves as an excitation donor molecule. 8-hydroxy-1,3,6-pyrenetrisulfonate (HPTS) chromophore in the deprotonated form serves as the acceptor of the excitation. In order to deprotonate the HPTS molecule pH of the water component of the solution was increased to 12.6. The change in pH did not appreciably affect the temperature of the phase transition of the water/2,6-lutidine mixture. HPTS molecule is fully ionized and highly charged. Consequently, it has a high affinity for the polar component of the mixture – water. Coumarine480 is a polar uncharged molecule, the competition of the electrostatic interaction and the energy, required to deform the hydrogen-bonded network of water forces the Coumarine480 molecule into the non-polar, 2,6-lutidine rich phase. The kinetics of the donor fluorescence decay as a function of temperature are shown on \ref{fig:id_expt}(b), the suppression of the fluorescence below the phase transition temperature of 307 degrees Kelvin is dramatic. The excitation transfer distance for a HPTS(deprotonate)-Coumarin480 pair of dye molecules was  measured to be 3.3nm. Absorption and emission properties of dye molecules were not altered in uniform and separated phases of the mixture. In the absence of the acceptor, the donor fluorescent lifetime did not change with the phase separation of the mixture and was measured to be 4.7ns.

 In order to instantaneously trigger the process of the phase transition we have chosen the system that undergoes the phase transition into an ordered state once the temperature is increased – the water/2,6-lutidine mixture (inverted phase diagram). It allows us to use a laser temperature jump to initiate the phase transition, as rapid as the vibrational relaxation of the molecules down to the picosecond time scale. Water/2,6-lutidine mixture was chosen at the critical composition with a lower consolute point at $T_c=$307 degrees Kelvin.

Figure \ref{fig:exp_sys} shows the experimental system employed for the temperature jump pump-excitation transfer probe experiment. Mid-IR temperature jump pulse was generated by KTA (KTiOAsO4), based optical parametric amplifier to produce tunable mid-IR radiation in the wavelength range 2.5-4 microns \cite{deak1997high, nisoli1994highly}. The mid-IR laser excitation pulse is sent to the mixture tuned to the mid-IR peak of water absorption at 3 microns. The transmission of the temperature jump beam through 5-micron thick sample at this wavelength was 50 percent. Temperature jump pulse energies were 3.1, 4.6, and 5.6 microjoules. With a heat capacity of water/2,6-lutidine mixture of 3.9 Joules/g*K at room temperature \cite{mirzaev2006critical}, these temperature jump pulse energies provide temperature jump magnitudes of 22, 32, and 41 degrees Kelvin. The initial temperautre of the mixture was the ambient temperature of the lab 295 Kelvin (22 degrees Centigrade). The long wavelength pass (LP) filter with a cut-off wavelength of 2.5 microns was employed to select the mid-IR wavelengths. 
 The visible (probe) portion of the experiment is based on the semiconductor diode laser  (Coherent, model CUBE 405-50C),  providing 50mW of power at 405nm, near the peak of the donor absorption. The visible and infrared beams were combined and co-aliened at the surface of LP filter which is transparent for the mid-IR wavelength and is reflective for 405nm light. Combined mid-IR and visible beams were focused by CaF2 lens with a focal distance of 2.5cm on the sample. The sample was rotated in order to avoid repeated exposure to mid-IR radiation. The 405nm radiation induces fluorescence in the sample which is collected by the lens of 5cm focal distance. In order to collect a large amount of fluorescence a large diameter (2-inch) lens was used. The center of the collecting lens was blocked, in order to prevent the probe beam from reaching the photomultiplier tube. The fluorescence, emitted by the sample was sent through the narrow band pass (10nm wide) filter centered at 440nm, towards the blue portion of the donor molecule emission. After the filter, the fluorescence was detected by the photomultiplier tube. The signal from the multiplier tube was digitized and stored in the computer. The mid-IR temperature jump pulse was modulated with a chopper, at the frequency of 500Hz. The fluorescence signal was digitized together with the information about the state of the mid-IR beam (present or absent). The change in the fluorescence signal due to a temperature jump pulse was extracted by lock-in amplification.
The beam sizes of the temperature jump beam and the fluorescence excitation beam were determined by sliding the razor blade across beams and measuring the transmitted intensity.
The diameter of the mid-IR beam was determined to be equal to 70 microns at full width ${1/e}$ intensity FW${1/e}$M. The 405nm fluorescence excitation beam has a diameter of 40 microns. The fluorescence excitation beam is well within the temperature jump beam.
	A spinning sample cell was used to expose the fresh spot for every temperature jump pulse. The sample was confined between CaF2 plates two inches in diameter. The spacing between CaF2 plates was 5 microns. The magnitude of the temperature jump was controlled by a combination of the half-wave plate and the mid-IR polarizer. Such a combination allows one to change the magnitude of the temperature jump without affecting the spatial overlap with a probe beam.

\begin{figure}
\includegraphics[width=0.4\textwidth]{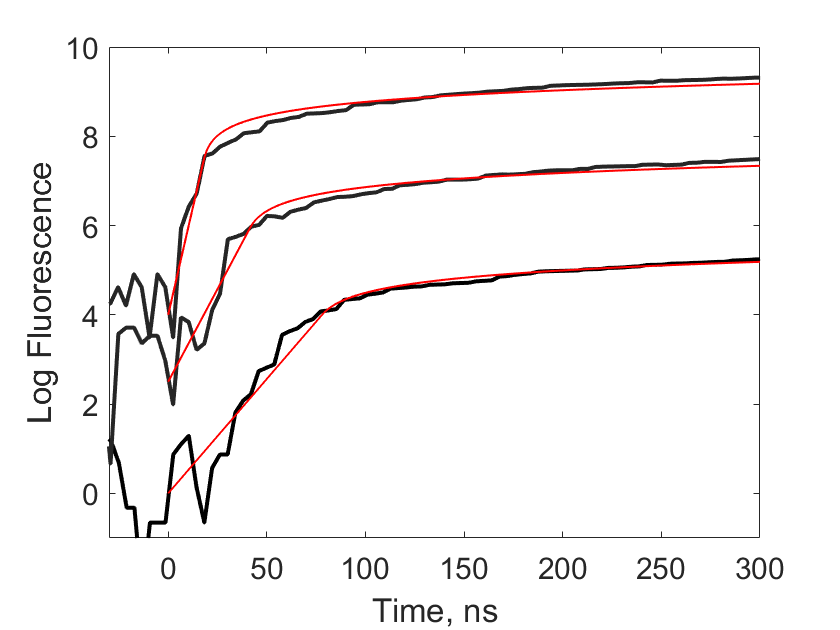}
\caption{\label{fig:kinetics} Kinetics of the donor fluorescence log scale after the initiation of the phase separation. The acceptor concentration is 50mM. Magnitudes of temperature jump: 40.5, 32.2, 22.1 (top to bottom). Fit underlines the exponential growth region of phase separation and transition to $t^{1/3}$ regime. Rates of phase separation: 0.19 $ns^{-1}$, 0.08 $ns^{-1}$, 0.05 $ns^{-1}$ . Transition time: {18.1 $ns$, 41.4 $ns$, 18.1 $ns$}}
\end{figure}

 Figure \ref{fig:kinetics} shows the kinetics of the fluorescence after the excitation with a temperature jump laser pulse. The vertical scale is logarithmic. Traces reflect the change in the kinetics for different magnitudes of temperature jumps. The initial temperature of the mixture was 293 degrees Kelvin. The following approximation form of the fluorescence kinetics was utilized: 
 \begin{equation}
s(t)= \left\{ \begin{array}{cl}
exp(t⁄\tau_{e}) & : \ t <\tau_c \\
A\cdot(t-(\tau_{c}-\tau_{e}/3))^{1/3}  & : \ t \geq \tau_{c}
\end{array} \right.
\end{equation}
where $A=(\tau_{e}/3)^{-1/3}exp(\tau_{e}⁄\tau_{e})$, $\tau_{e}$ is the exponential growth time scale, $\tau_{c}$ is the transition time from the exponential to ${t^{1/3}}$ coarsening regime of phase separation.

\begin{figure}
\includegraphics[width=0.4\textwidth]{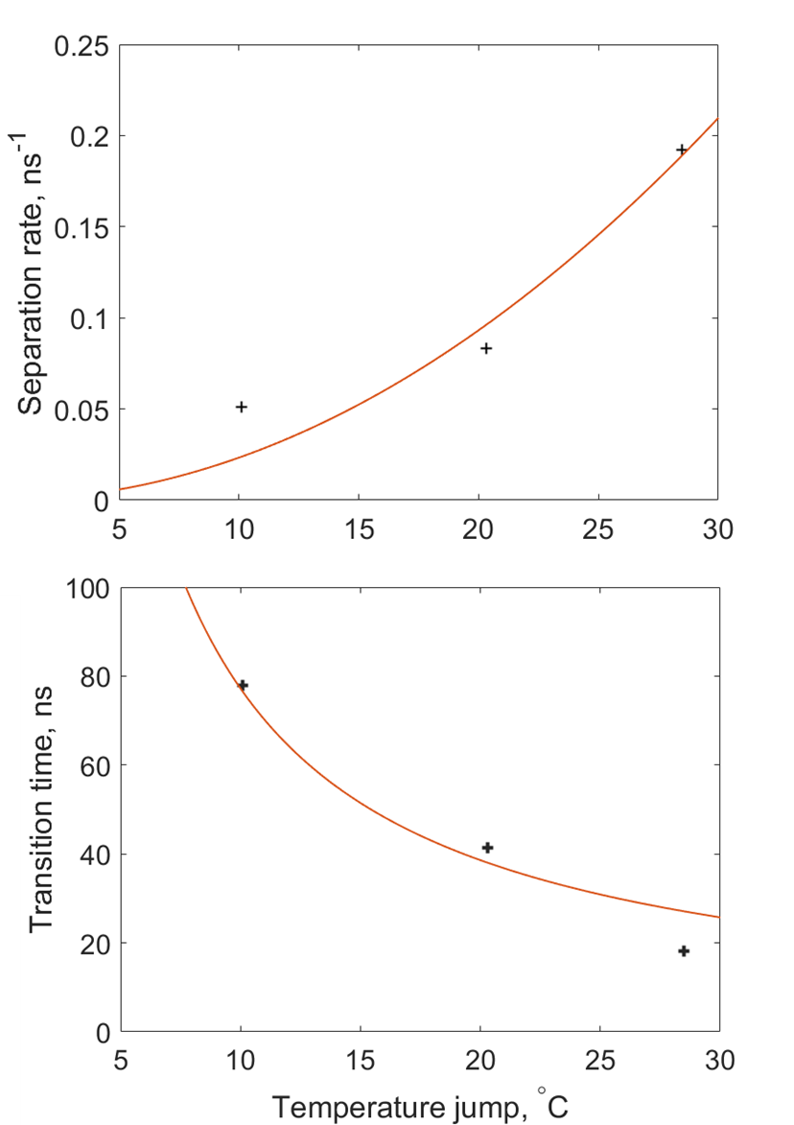}
\caption{\label{fig:rate} Rate of the exponential stage of the phase separation as a function of the temperature excursion above the $T_c$.}
\end{figure}

Figure \ref{fig:rate} shows the phase separation rate (in inverse nanoseconds) as a function of the temperature jump magnitude. 

According to Cahn \cite{cahn1961spinodal}, the relation between the time scale of spinodal decomposition and the phase separation distance is given by 

\begin{equation}
 \lambda\simeq \pi\sqrt{8D\tau}   
\end{equation}

According to \cite{clunie1999interdiffusion}, the interdiffusion coefficient for components of water/2,6-lutidine mixture at critical composition is $1.1\times 10^{-6} cm^2/s$ and is weakly dependent on the temperature away from a critical point. For temperature jumps utilized in the experiment, the space scales of phase separation are 13.0 nm, 10.2 nm, and 6.7 nm. The space scale of phase separation is larger than the hydrodynamic radius of HPTS and Coumarine 450 equal to 0.5nm.

After a temperature jump the system finds itself in an unstable state and starts a diffusion towards new spatial organization \cite{cahn1961spinodal}, the growth rate of composition fluctuations at the maximum instability is given by

\begin{equation} \label{highest_rate}
k_{max}=\frac{3}{32}M\left(\left(\frac{\partial^2f}{\partial c^2}\right)_{c=c_{i}}\right)^2  
\end{equation}
where $f$ is the free energy, $M$ the empirical diffusion coefficient \cite{cahn1961spinodal}, $c_{i}$ is the initial composition.
When the temperature rises above the critical $T_c$ the second derivative of free energy with respect to the composition changes its sign to make the initial composition unstable. Thus, the temperature dependence of the free energy second derivative can be represented as $\left(\frac{\partial^2f}{\partial c^2}\right)_{c=c_i} \sim \alpha\left(T-T_c\right)$. Figure \ref{fig:rate} top shows the experimentally measured dependence of the phase transformation rate on the difference between the final and critical temperatures, $\Delta T$. The approximating curve is $k=\alpha \cdot (\Delta T)^{2}$, $\alpha=2.3 \cdot 10^{-4}  ns^{-1}/{K^{2}}$.  Figure \ref{fig:rate} bottom panel shows the transition time to the polynomial growth, which can be approximated as $\tau_{c}=780/\Delta T$

Thus, the application of FRET allows one to obtain the structural information during the phase transformation on a nanometer space scale with nanosecond resolution. If instead of CW probe beam one applies the stream of femtosecond pulses, provided by the oscillator, then not only the total amount of fluorescens but the kinetics of its decay can be measured, providing additional structural information.

The rate of the exponential growth of phase transition was an order of magnitude larger than predicted by \cite{hazra2016nanoscale, hazra2019three}. The rate of the exponential stage of the phase transition was increasing and the transition to the later stage of spinodal decomposition was occurring earlier in agreement with \cite{hazra2016nanoscale, hazra2019three}.

Thus, we have demonstrated a route for measurement of the earliest stages of phase separation, that allows one to study the geometry of the nucleation site through careful interpretation of FRET excitation decay kinetics. A similar approach can be applied in a broad range of chemical and biological systems \cite{banani2017biomolecular, shin2017liquid, yao2008effects, kim2007phase}.
	
\begin{acknowledgments}
I would like to acknowledge the support and valuable discussion with Michael D. Fayer Professor of Chemistry, Stanford University in carrying out the experimental work.
\end{acknowledgments}

\nocite{*}

\bibliography{apssamp}

\providecommand{\noopsort}[1]{}\providecommand{\singleletter}[1]{#1}%
\begin{thebibliography}{23}%
\makeatletter
\providecommand \@ifxundefined [1]{%
 \@ifx{#1\undefined}
}%
\providecommand \@ifnum [1]{%
 \ifnum #1\expandafter \@firstoftwo
 \else \expandafter \@secondoftwo
 \fi
}%
\providecommand \@ifx [1]{%
 \ifx #1\expandafter \@firstoftwo
 \else \expandafter \@secondoftwo
 \fi
}%
\providecommand \natexlab [1]{#1}%
\providecommand \enquote  [1]{``#1''}%
\providecommand \bibnamefont  [1]{#1}%
\providecommand \bibfnamefont [1]{#1}%
\providecommand \citenamefont [1]{#1}%
\providecommand \href@noop [0]{\@secondoftwo}%
\providecommand \href [0]{\begingroup \@sanitize@url \@href}%
\providecommand \@href[1]{\@@startlink{#1}\@@href}%
\providecommand \@@href[1]{\endgroup#1\@@endlink}%
\providecommand \@sanitize@url [0]{\catcode `\\12\catcode `\$12\catcode
  `\&12\catcode `\#12\catcode `\^12\catcode `\_12\catcode `\%12\relax}%
\providecommand \@@startlink[1]{}%
\providecommand \@@endlink[0]{}%
\providecommand \url  [0]{\begingroup\@sanitize@url \@url }%
\providecommand \@url [1]{\endgroup\@href {#1}{\urlprefix }}%
\providecommand \urlprefix  [0]{URL }%
\providecommand \Eprint [0]{\href }%
\providecommand \doibase [0]{https://doi.org/}%
\providecommand \selectlanguage [0]{\@gobble}%
\providecommand \bibinfo  [0]{\@secondoftwo}%
\providecommand \bibfield  [0]{\@secondoftwo}%
\providecommand \translation [1]{[#1]}%
\providecommand \BibitemOpen [0]{}%
\providecommand \bibitemStop [0]{}%
\providecommand \bibitemNoStop [0]{.\EOS\space}%
\providecommand \EOS [0]{\spacefactor3000\relax}%
\providecommand \BibitemShut  [1]{\csname bibitem#1\endcsname}%
\let\auto@bib@innerbib\@empty
\bibitem [{\citenamefont {Rosenzweig}\ \emph {et~al.}(2017)\citenamefont
  {Rosenzweig}, \citenamefont {Xu}, \citenamefont {Cuellar}, \citenamefont
  {Martinez-Sanchez}, \citenamefont {Schaffer}, \citenamefont {Strauss},
  \citenamefont {Cartwright}, \citenamefont {Ronceray}, \citenamefont
  {Plitzko}, \citenamefont {F{\"o}rster} \emph
  {et~al.}}]{rosenzweig2017eukaryotic}%
  \BibitemOpen
  \bibfield  {author} {\bibinfo {author} {\bibfnamefont {E.~S.~F.}\
  \bibnamefont {Rosenzweig}}, \bibinfo {author} {\bibfnamefont
  {B.}~\bibnamefont {Xu}}, \bibinfo {author} {\bibfnamefont {L.~K.}\
  \bibnamefont {Cuellar}}, \bibinfo {author} {\bibfnamefont {A.}~\bibnamefont
  {Martinez-Sanchez}}, \bibinfo {author} {\bibfnamefont {M.}~\bibnamefont
  {Schaffer}}, \bibinfo {author} {\bibfnamefont {M.}~\bibnamefont {Strauss}},
  \bibinfo {author} {\bibfnamefont {H.~N.}\ \bibnamefont {Cartwright}},
  \bibinfo {author} {\bibfnamefont {P.}~\bibnamefont {Ronceray}}, \bibinfo
  {author} {\bibfnamefont {J.~M.}\ \bibnamefont {Plitzko}}, \bibinfo {author}
  {\bibfnamefont {F.}~\bibnamefont {F{\"o}rster}}, \emph {et~al.},\ }\bibfield
  {title} {\bibinfo {title} {The eukaryotic co2-concentrating organelle is
  liquid-like and exhibits dynamic reorganization},\ }\href@noop {} {\bibfield
  {journal} {\bibinfo  {journal} {Cell}\ }\textbf {\bibinfo {volume} {171}},\
  \bibinfo {pages} {148} (\bibinfo {year} {2017})}\BibitemShut {NoStop}%
\bibitem [{\citenamefont {Bates}\ and\ \citenamefont
  {Wiltzius}(1989)}]{bates1989spinodal}%
  \BibitemOpen
  \bibfield  {author} {\bibinfo {author} {\bibfnamefont {F.~S.}\ \bibnamefont
  {Bates}}\ and\ \bibinfo {author} {\bibfnamefont {P.}~\bibnamefont
  {Wiltzius}},\ }\bibfield  {title} {\bibinfo {title} {Spinodal decomposition
  of a symmetric critical mixture of deuterated and protonated polymer},\
  }\href@noop {} {\bibfield  {journal} {\bibinfo  {journal} {The Journal of
  chemical physics}\ }\textbf {\bibinfo {volume} {91}},\ \bibinfo {pages}
  {3258} (\bibinfo {year} {1989})}\BibitemShut {NoStop}%
\bibitem [{\citenamefont {Tran-Cong}\ \emph {et~al.}(1997)\citenamefont
  {Tran-Cong}, \citenamefont {Ohta},\ and\ \citenamefont
  {Urakawa}}]{tran1997soft}%
  \BibitemOpen
  \bibfield  {author} {\bibinfo {author} {\bibfnamefont {Q.}~\bibnamefont
  {Tran-Cong}}, \bibinfo {author} {\bibfnamefont {T.}~\bibnamefont {Ohta}},\
  and\ \bibinfo {author} {\bibfnamefont {O.}~\bibnamefont {Urakawa}},\
  }\bibfield  {title} {\bibinfo {title} {Soft-mode suppression in the phase
  separation of binary polymer mixtures driven by a reversible chemical
  reaction},\ }\href@noop {} {\bibfield  {journal} {\bibinfo  {journal}
  {Physical Review E}\ }\textbf {\bibinfo {volume} {56}},\ \bibinfo {pages}
  {R59} (\bibinfo {year} {1997})}\BibitemShut {NoStop}%
\bibitem [{\citenamefont {Mallamace}\ \emph {et~al.}(1995)\citenamefont
  {Mallamace}, \citenamefont {Micali}, \citenamefont {Trusso},\ and\
  \citenamefont {Chen}}]{mallamace1995spinodal}%
  \BibitemOpen
  \bibfield  {author} {\bibinfo {author} {\bibfnamefont {F.}~\bibnamefont
  {Mallamace}}, \bibinfo {author} {\bibfnamefont {N.}~\bibnamefont {Micali}},
  \bibinfo {author} {\bibfnamefont {S.}~\bibnamefont {Trusso}},\ and\ \bibinfo
  {author} {\bibfnamefont {S.}~\bibnamefont {Chen}},\ }\bibfield  {title}
  {\bibinfo {title} {Spinodal decomposition of a three-component water-in-oil
  microemulsion system},\ }\href@noop {} {\bibfield  {journal} {\bibinfo
  {journal} {Physical Review E}\ }\textbf {\bibinfo {volume} {51}},\ \bibinfo
  {pages} {5818} (\bibinfo {year} {1995})}\BibitemShut {NoStop}%
\bibitem [{\citenamefont {Kubota}\ and\ \citenamefont
  {Kuwahara}(1992)}]{kubota1992spinodal}%
  \BibitemOpen
  \bibfield  {author} {\bibinfo {author} {\bibfnamefont {K.}~\bibnamefont
  {Kubota}}\ and\ \bibinfo {author} {\bibfnamefont {N.}~\bibnamefont
  {Kuwahara}},\ }\bibfield  {title} {\bibinfo {title} {Spinodal decomposition
  in a binary mixture},\ }\href@noop {} {\bibfield  {journal} {\bibinfo
  {journal} {Physical review letters}\ }\textbf {\bibinfo {volume} {68}},\
  \bibinfo {pages} {197} (\bibinfo {year} {1992})}\BibitemShut {NoStop}%
\bibitem [{\citenamefont {Kubota}\ \emph {et~al.}(1992)\citenamefont {Kubota},
  \citenamefont {Kuwahara}, \citenamefont {Eda},\ and\ \citenamefont
  {Sakazume}}]{kubota1992spinodal_PRA}%
  \BibitemOpen
  \bibfield  {author} {\bibinfo {author} {\bibfnamefont {K.}~\bibnamefont
  {Kubota}}, \bibinfo {author} {\bibfnamefont {N.}~\bibnamefont {Kuwahara}},
  \bibinfo {author} {\bibfnamefont {H.}~\bibnamefont {Eda}},\ and\ \bibinfo
  {author} {\bibfnamefont {M.}~\bibnamefont {Sakazume}},\ }\bibfield  {title}
  {\bibinfo {title} {Spinodal decomposition in a critical isobutyric acid and
  water mixture},\ }\href@noop {} {\bibfield  {journal} {\bibinfo  {journal}
  {Physical Review A}\ }\textbf {\bibinfo {volume} {45}},\ \bibinfo {pages}
  {R3377} (\bibinfo {year} {1992})}\BibitemShut {NoStop}%
\bibitem [{\citenamefont {Chou}\ and\ \citenamefont
  {Goldburg}(1979)}]{chou1979phase}%
  \BibitemOpen
  \bibfield  {author} {\bibinfo {author} {\bibfnamefont {Y.-C.}\ \bibnamefont
  {Chou}}\ and\ \bibinfo {author} {\bibfnamefont {W.~I.}\ \bibnamefont
  {Goldburg}},\ }\bibfield  {title} {\bibinfo {title} {Phase separation and
  coalescence in critically quenched isobutyric-acid—water and 2,
  6-lutidine—water mixtures},\ }\href@noop {} {\bibfield  {journal} {\bibinfo
   {journal} {Physical Review A}\ }\textbf {\bibinfo {volume} {20}},\ \bibinfo
  {pages} {2105} (\bibinfo {year} {1979})}\BibitemShut {NoStop}%
\bibitem [{\citenamefont {Huang}\ \emph {et~al.}(1974)\citenamefont {Huang},
  \citenamefont {Goldburg},\ and\ \citenamefont {Bjerkaas}}]{huang1974study}%
  \BibitemOpen
  \bibfield  {author} {\bibinfo {author} {\bibfnamefont {J.~S.}\ \bibnamefont
  {Huang}}, \bibinfo {author} {\bibfnamefont {W.~I.}\ \bibnamefont
  {Goldburg}},\ and\ \bibinfo {author} {\bibfnamefont {A.~W.}\ \bibnamefont
  {Bjerkaas}},\ }\bibfield  {title} {\bibinfo {title} {Study of phase
  separation in a critical binary liquid mixture: Spinodal decomposition},\
  }\href@noop {} {\bibfield  {journal} {\bibinfo  {journal} {Physical Review
  Letters}\ }\textbf {\bibinfo {volume} {32}},\ \bibinfo {pages} {921}
  (\bibinfo {year} {1974})}\BibitemShut {NoStop}%
\bibitem [{\citenamefont {Pekker}\ and\ \citenamefont
  {Shneider}(2017)}]{pekker2017initial}%
  \BibitemOpen
  \bibfield  {author} {\bibinfo {author} {\bibfnamefont {M.}~\bibnamefont
  {Pekker}}\ and\ \bibinfo {author} {\bibfnamefont {M.}~\bibnamefont
  {Shneider}},\ }\bibfield  {title} {\bibinfo {title} {Initial stage of
  cavitation in liquids and its observation by rayleigh scattering},\
  }\href@noop {} {\bibfield  {journal} {\bibinfo  {journal} {Fluid Dynamics
  Research}\ }\textbf {\bibinfo {volume} {49}},\ \bibinfo {pages} {035503}
  (\bibinfo {year} {2017})}\BibitemShut {NoStop}%
\bibitem [{\citenamefont {De{\`a}k}\ \emph {et~al.}(1997)\citenamefont
  {De{\`a}k}, \citenamefont {Iwaki},\ and\ \citenamefont
  {Dlott}}]{deak1997high}%
  \BibitemOpen
  \bibfield  {author} {\bibinfo {author} {\bibfnamefont {J.~C.}\ \bibnamefont
  {De{\`a}k}}, \bibinfo {author} {\bibfnamefont {L.~K.}\ \bibnamefont
  {Iwaki}},\ and\ \bibinfo {author} {\bibfnamefont {D.~D.}\ \bibnamefont
  {Dlott}},\ }\bibfield  {title} {\bibinfo {title} {High-power picosecond
  mid-infrared optical parametric amplifier for infrared raman spectroscopy},\
  }\href@noop {} {\bibfield  {journal} {\bibinfo  {journal} {Optics letters}\
  }\textbf {\bibinfo {volume} {22}},\ \bibinfo {pages} {1796} (\bibinfo {year}
  {1997})}\BibitemShut {NoStop}%
\bibitem [{\citenamefont {Nisoli}\ \emph {et~al.}(1994)\citenamefont {Nisoli},
  \citenamefont {De~Silvestri}, \citenamefont {Magni}, \citenamefont {Svelto},
  \citenamefont {Danielius}, \citenamefont {Piskarskas}, \citenamefont
  {Valiulis},\ and\ \citenamefont {Varanavicius}}]{nisoli1994highly}%
  \BibitemOpen
  \bibfield  {author} {\bibinfo {author} {\bibfnamefont {M.}~\bibnamefont
  {Nisoli}}, \bibinfo {author} {\bibfnamefont {S.}~\bibnamefont
  {De~Silvestri}}, \bibinfo {author} {\bibfnamefont {V.}~\bibnamefont {Magni}},
  \bibinfo {author} {\bibfnamefont {O.}~\bibnamefont {Svelto}}, \bibinfo
  {author} {\bibfnamefont {R.}~\bibnamefont {Danielius}}, \bibinfo {author}
  {\bibfnamefont {A.}~\bibnamefont {Piskarskas}}, \bibinfo {author}
  {\bibfnamefont {G.}~\bibnamefont {Valiulis}},\ and\ \bibinfo {author}
  {\bibfnamefont {A.}~\bibnamefont {Varanavicius}},\ }\bibfield  {title}
  {\bibinfo {title} {Highly efficient parametric conversion of femtosecond ti:
  sapphire laser pulses at 1 khz},\ }\href@noop {} {\bibfield  {journal}
  {\bibinfo  {journal} {Optics letters}\ }\textbf {\bibinfo {volume} {19}},\
  \bibinfo {pages} {1973} (\bibinfo {year} {1994})}\BibitemShut {NoStop}%
\bibitem [{\citenamefont {Mirzaev}\ \emph {et~al.}(2006)\citenamefont
  {Mirzaev}, \citenamefont {Behrends}, \citenamefont {Heimburg}, \citenamefont
  {Haller},\ and\ \citenamefont {Kaatze}}]{mirzaev2006critical}%
  \BibitemOpen
  \bibfield  {author} {\bibinfo {author} {\bibfnamefont {S.~Z.}\ \bibnamefont
  {Mirzaev}}, \bibinfo {author} {\bibfnamefont {R.}~\bibnamefont {Behrends}},
  \bibinfo {author} {\bibfnamefont {T.}~\bibnamefont {Heimburg}}, \bibinfo
  {author} {\bibfnamefont {J.}~\bibnamefont {Haller}},\ and\ \bibinfo {author}
  {\bibfnamefont {U.}~\bibnamefont {Kaatze}},\ }\bibfield  {title} {\bibinfo
  {title} {Critical behavior of 2, 6-dimethylpyridine-water: Measurements of
  specific heat, dynamic light scattering, and shear viscosity},\ }\href@noop
  {} {\bibfield  {journal} {\bibinfo  {journal} {The Journal of chemical
  physics}\ }\textbf {\bibinfo {volume} {124}} (\bibinfo {year}
  {2006})}\BibitemShut {NoStop}%
\bibitem [{\citenamefont {Cahn}(1961)}]{cahn1961spinodal}%
  \BibitemOpen
  \bibfield  {author} {\bibinfo {author} {\bibfnamefont {J.~W.}\ \bibnamefont
  {Cahn}},\ }\bibfield  {title} {\bibinfo {title} {On spinodal decomposition},\
  }\href@noop {} {\bibfield  {journal} {\bibinfo  {journal} {Acta
  metallurgica}\ }\textbf {\bibinfo {volume} {9}},\ \bibinfo {pages} {795}
  (\bibinfo {year} {1961})}\BibitemShut {NoStop}%
\bibitem [{\citenamefont {Clunie}\ and\ \citenamefont
  {Baird}(1999)}]{clunie1999interdiffusion}%
  \BibitemOpen
  \bibfield  {author} {\bibinfo {author} {\bibfnamefont {J.~C.}\ \bibnamefont
  {Clunie}}\ and\ \bibinfo {author} {\bibfnamefont {J.~K.}\ \bibnamefont
  {Baird}},\ }\bibfield  {title} {\bibinfo {title} {Interdiffusion coefficient
  and dynamic viscosity for the mixture 2, 6-lutidine+ water near the lower
  consolute point},\ }\href@noop {} {\bibfield  {journal} {\bibinfo  {journal}
  {Physics and Chemistry of Liquids}\ }\textbf {\bibinfo {volume} {37}},\
  \bibinfo {pages} {357} (\bibinfo {year} {1999})}\BibitemShut {NoStop}%
\bibitem [{\citenamefont {Hazra}\ \emph {et~al.}(2016)\citenamefont {Hazra},
  \citenamefont {Sarkar},\ and\ \citenamefont {Bagchi}}]{hazra2016nanoscale}%
  \BibitemOpen
  \bibfield  {author} {\bibinfo {author} {\bibfnamefont {M.~K.}\ \bibnamefont
  {Hazra}}, \bibinfo {author} {\bibfnamefont {S.}~\bibnamefont {Sarkar}},\ and\
  \bibinfo {author} {\bibfnamefont {B.}~\bibnamefont {Bagchi}},\ }\bibfield
  {title} {\bibinfo {title} {Nanoscale heterogeneous phase separation kinetics
  in binary mixtures: Multistage dynamics},\ }\href@noop {} {\bibfield
  {journal} {\bibinfo  {journal} {arXiv preprint arXiv:1604.02873}\ } (\bibinfo
  {year} {2016})}\BibitemShut {NoStop}%
\bibitem [{\citenamefont {Hazra}\ \emph {et~al.}(2019)\citenamefont {Hazra},
  \citenamefont {Sarkar},\ and\ \citenamefont {Bagchi}}]{hazra2019three}%
  \BibitemOpen
  \bibfield  {author} {\bibinfo {author} {\bibfnamefont {M.~K.}\ \bibnamefont
  {Hazra}}, \bibinfo {author} {\bibfnamefont {S.}~\bibnamefont {Sarkar}},\ and\
  \bibinfo {author} {\bibfnamefont {B.}~\bibnamefont {Bagchi}},\ }\bibfield
  {title} {\bibinfo {title} {Three-stage phase separation kinetics in a model
  liquid binary mixture: A computational study},\ }\href@noop {} {\bibfield
  {journal} {\bibinfo  {journal} {The Journal of Chemical Physics}\ }\textbf
  {\bibinfo {volume} {150}} (\bibinfo {year} {2019})}\BibitemShut {NoStop}%
\bibitem [{\citenamefont {Banani}\ \emph {et~al.}(2017)\citenamefont {Banani},
  \citenamefont {Lee}, \citenamefont {Hyman},\ and\ \citenamefont
  {Rosen}}]{banani2017biomolecular}%
  \BibitemOpen
  \bibfield  {author} {\bibinfo {author} {\bibfnamefont {S.~F.}\ \bibnamefont
  {Banani}}, \bibinfo {author} {\bibfnamefont {H.~O.}\ \bibnamefont {Lee}},
  \bibinfo {author} {\bibfnamefont {A.~A.}\ \bibnamefont {Hyman}},\ and\
  \bibinfo {author} {\bibfnamefont {M.~K.}\ \bibnamefont {Rosen}},\ }\bibfield
  {title} {\bibinfo {title} {Biomolecular condensates: organizers of cellular
  biochemistry},\ }\href@noop {} {\bibfield  {journal} {\bibinfo  {journal}
  {Nature reviews Molecular cell biology}\ }\textbf {\bibinfo {volume} {18}},\
  \bibinfo {pages} {285} (\bibinfo {year} {2017})}\BibitemShut {NoStop}%
\bibitem [{\citenamefont {Shin}\ and\ \citenamefont
  {Brangwynne}(2017)}]{shin2017liquid}%
  \BibitemOpen
  \bibfield  {author} {\bibinfo {author} {\bibfnamefont {Y.}~\bibnamefont
  {Shin}}\ and\ \bibinfo {author} {\bibfnamefont {C.~P.}\ \bibnamefont
  {Brangwynne}},\ }\bibfield  {title} {\bibinfo {title} {Liquid phase
  condensation in cell physiology and disease},\ }\href@noop {} {\bibfield
  {journal} {\bibinfo  {journal} {Science}\ }\textbf {\bibinfo {volume}
  {357}},\ \bibinfo {pages} {eaaf4382} (\bibinfo {year} {2017})}\BibitemShut
  {NoStop}%
\bibitem [{\citenamefont {Yao}\ \emph {et~al.}(2008)\citenamefont {Yao},
  \citenamefont {Hou}, \citenamefont {Xu}, \citenamefont {Li},\ and\
  \citenamefont {Yang}}]{yao2008effects}%
  \BibitemOpen
  \bibfield  {author} {\bibinfo {author} {\bibfnamefont {Y.}~\bibnamefont
  {Yao}}, \bibinfo {author} {\bibfnamefont {J.}~\bibnamefont {Hou}}, \bibinfo
  {author} {\bibfnamefont {Z.}~\bibnamefont {Xu}}, \bibinfo {author}
  {\bibfnamefont {G.}~\bibnamefont {Li}},\ and\ \bibinfo {author}
  {\bibfnamefont {Y.}~\bibnamefont {Yang}},\ }\bibfield  {title} {\bibinfo
  {title} {Effects of solvent mixtures on the nanoscale phase separation in
  polymer solar cells},\ }\href@noop {} {\bibfield  {journal} {\bibinfo
  {journal} {Advanced Functional Materials}\ }\textbf {\bibinfo {volume}
  {18}},\ \bibinfo {pages} {1783} (\bibinfo {year} {2008})}\BibitemShut
  {NoStop}%
\bibitem [{\citenamefont {Kim}\ \emph {et~al.}(2007)\citenamefont {Kim},
  \citenamefont {Choi}, \citenamefont {Hong},\ and\ \citenamefont
  {Kim}}]{kim2007phase}%
  \BibitemOpen
  \bibfield  {author} {\bibinfo {author} {\bibfnamefont {D.~H.}\ \bibnamefont
  {Kim}}, \bibinfo {author} {\bibfnamefont {J.}~\bibnamefont {Choi}}, \bibinfo
  {author} {\bibfnamefont {Y.~T.}\ \bibnamefont {Hong}},\ and\ \bibinfo
  {author} {\bibfnamefont {S.~C.}\ \bibnamefont {Kim}},\ }\bibfield  {title}
  {\bibinfo {title} {Phase separation and morphology control of polymer blend
  membranes of sulfonated and nonsulfonated polysulfones for direct methanol
  fuel cell application},\ }\href@noop {} {\bibfield  {journal} {\bibinfo
  {journal} {Journal of membrane science}\ }\textbf {\bibinfo {volume} {299}},\
  \bibinfo {pages} {19} (\bibinfo {year} {2007})}\BibitemShut {NoStop}%
\bibitem [{\citenamefont {Takamuku}\ \emph {et~al.}(2001)\citenamefont
  {Takamuku}, \citenamefont {Yamaguchi}, \citenamefont {Matsuo}, \citenamefont
  {Tabata}, \citenamefont {Kumamoto}, \citenamefont {Nishimoto}, \citenamefont
  {Yoshida}, \citenamefont {Yamaguchi}, \citenamefont {Nagao}, \citenamefont
  {Otomo} \emph {et~al.}}]{takamuku2001large}%
  \BibitemOpen
  \bibfield  {author} {\bibinfo {author} {\bibfnamefont {T.}~\bibnamefont
  {Takamuku}}, \bibinfo {author} {\bibfnamefont {A.}~\bibnamefont {Yamaguchi}},
  \bibinfo {author} {\bibfnamefont {D.}~\bibnamefont {Matsuo}}, \bibinfo
  {author} {\bibfnamefont {M.}~\bibnamefont {Tabata}}, \bibinfo {author}
  {\bibfnamefont {M.}~\bibnamefont {Kumamoto}}, \bibinfo {author}
  {\bibfnamefont {J.}~\bibnamefont {Nishimoto}}, \bibinfo {author}
  {\bibfnamefont {K.}~\bibnamefont {Yoshida}}, \bibinfo {author} {\bibfnamefont
  {T.}~\bibnamefont {Yamaguchi}}, \bibinfo {author} {\bibfnamefont
  {M.}~\bibnamefont {Nagao}}, \bibinfo {author} {\bibfnamefont
  {T.}~\bibnamefont {Otomo}}, \emph {et~al.},\ }\bibfield  {title} {\bibinfo
  {title} {Large-angle x-ray scattering and small-angle neutron scattering
  study on phase separation of acetonitrile- water mixtures by addition of
  nacl},\ }\href@noop {} {\bibfield  {journal} {\bibinfo  {journal} {The
  Journal of Physical Chemistry B}\ }\textbf {\bibinfo {volume} {105}},\
  \bibinfo {pages} {6236} (\bibinfo {year} {2001})}\BibitemShut {NoStop}%
\bibitem [{\citenamefont {Yoshida}\ \emph {et~al.}(2000)\citenamefont
  {Yoshida}, \citenamefont {Misawa}, \citenamefont {Maruyama}, \citenamefont
  {Imai},\ and\ \citenamefont {Furusaka}}]{yoshida2000small}%
  \BibitemOpen
  \bibfield  {author} {\bibinfo {author} {\bibfnamefont {K.}~\bibnamefont
  {Yoshida}}, \bibinfo {author} {\bibfnamefont {M.}~\bibnamefont {Misawa}},
  \bibinfo {author} {\bibfnamefont {K.}~\bibnamefont {Maruyama}}, \bibinfo
  {author} {\bibfnamefont {M.}~\bibnamefont {Imai}},\ and\ \bibinfo {author}
  {\bibfnamefont {M.}~\bibnamefont {Furusaka}},\ }\bibfield  {title} {\bibinfo
  {title} {Small angle neutron scattering study on the salt-induced phase
  separation of 1-propanol aqueous solution},\ }\href@noop {} {\bibfield
  {journal} {\bibinfo  {journal} {The journal of chemical physics}\ }\textbf
  {\bibinfo {volume} {113}},\ \bibinfo {pages} {2343} (\bibinfo {year}
  {2000})}\BibitemShut {NoStop}%
\bibitem [{\citenamefont {Yano}(2002)}]{yano2002surface}%
  \BibitemOpen
  \bibfield  {author} {\bibinfo {author} {\bibfnamefont {Y.~F.}\ \bibnamefont
  {Yano}},\ }\bibfield  {title} {\bibinfo {title} {Surface structure of aqueous
  2-butoxyethanol mixtures studied by x-ray reflection},\ }\href@noop {}
  {\bibfield  {journal} {\bibinfo  {journal} {The Journal of chemical physics}\
  }\textbf {\bibinfo {volume} {116}},\ \bibinfo {pages} {8093} (\bibinfo {year}
  {2002})}\BibitemShut {NoStop}%
\end{thebibliography}%

\end{document}